\documentclass[12pt]{article}
\pdfoutput=1

\usepackage{amsmath,amssymb,graphicx}
\usepackage{epsf}
\usepackage{pstricks}
\usepackage{cite}
\usepackage{multirow}
\usepackage{array}
\usepackage{hyperref}

\newcommand{\beq}{\begin{eqnarray}}
\newcommand{\eeq}{\end{eqnarray}}
\def\ma{m_{h_1}}
\def\mb{m_{h_2}}
\def\maa{m_{a_1}}
\def\mhp{m_{H^+}}
\def\tb{\tan\beta}
\def\mueff{\mu_\text{eff}}
\def\kva{\kappa_{V,1}}
\def\kvb{\kappa_{V,2}}

\def\kta{\kappa_{t,1}}
\def\ktb{\kappa_{t,2}}
\def\kba{\kappa_{b,1}}
\def\kbb{\kappa_{b,2}}

\def\klb{\kappa_{\tau,2}}
\def\kfa{\kappa_{f,1}}
\def\kfb{\kappa_{f,2}}
\def\kgama{\kappa_{\gamma,1}}

\def\talpha{\tilde{\alpha}}
\def\sa{s_{\talpha}}
\def\ca{c_{\talpha}}
\def\A{\mathcal{A}}
\def\ggh{\text{ggh}}
\def\VBF{\text{VBF}}
\def\VH{\text{VH}}

\def\bsg{B\to X_s\gamma}
\def\bsmumu{B_s\to\bar\mu\mu}
\def\dmd{\Delta M_d}
\def\superiso{\texttt{SuperIso}}
\def\thdmc{\texttt{2HDMC}}
\def\hb{\texttt{HiggsBounds}}

\begin{document}
\begin{titlepage}
\begin{flushright}
LYCEN 2013-17 
\end{flushright}

\vskip.5cm
\begin{center}
{\huge \bf Searching for a lighter
Higgs: parametrisation and sample tests}

\vskip.1cm
\end{center}
\vskip1cm

\begin{center}
{\bf {Giacomo Cacciapaglia}, {Aldo Deandrea\footnote{also Institut Universitaire de France, 103 boulevard Saint-Michel, 
75005 Paris, France}}, {Guillaume Drieu La Rochelle} and {Jean-Baptiste Flament}}
\end{center}
\vskip 8pt

\begin{center}
{\it
Universit\'e de Lyon, F-69622 Lyon, France; Universit\'e Lyon 1, Villeurbanne;\\
CNRS/IN2P3, UMR5822, Institut de Physique Nucl\'eaire de Lyon\\
F-69622 Villeurbanne Cedex, France} \\
\vspace*{.5cm}
{\tt  g.cacciapaglia@ipnl.in2p3.fr, deandrea@ipnl.in2p3.fr, g.drieu-la-rochelle@ipnl.in2p3.fr, j-b.flament@ipnl.in2p3.fr}
\end{center}

\vskip2truecm

\begin{abstract}
\vskip 3pt
\noindent
The structure of the Higgs sector is a major issue in the quest of a detailed description of the electroweak interactions. Most of the 
effort is devoted to the study of the standard model--like Higgs boson at 126 GeV, however the experimental collaborations at the LHC are also 
searching for extra scalar particles whose presence may hint to an extended Higgs sector, typical of many extensions of the standard model.
We study a model independent parametrisation of a scalar particle lighter than the 126 GeV Higgs boson, which may be easily 
implemented in the ongoing searches by ATLAS and CMS. 
Indeed many effective Lagrangians/parametrisations used at present for the description of the Higgs sector implicitly assume that no 
light particles other than the standard model ones are present in the spectrum. We therefore introduce a parametrisation of a two 
scalars model, one corresponding to the 126 GeV Higgs boson and the other to a lighter scalar. After the introduction of such a tool, 
we consider two typical examples falling in this category,  in order to illustrate the use of our formalism: the two Higgs doublet model 
and the next to minimal supersymmetric standard model. Our results agree with the specific studies performed for these models. Furthermore, 
employing such a generic parametrisation allows us to translate the bounds to any model beyond the standard model falling into 
this class.
\end{abstract}

\end{titlepage}

\newpage

\section{Introduction}
With the increasingly accurate measurements of the coupling properties of the Higgs boson at the LHC, the whole phenomenology of 
electroweak physics is entering a new era. It is widely thought that, although the Higgs itself is no evidence for physics Beyond the 
Standard Model (BSM), there are still various points upon which New Physics would be required, as for instance in providing a dark 
matter candidate or by solving the naturalness issue. Within BSM theories, the coupling properties of the Higgs are generically 
not fixed, making the accurate measurements of those an important constraint to take into account. This constraint is particularly 
crucial in extensions of the Standard Model (SM) where new scalars are added. In this case, one expects the properties of the 
various scalars to be related, thus allowing a correlation between the determination of the couplings of the SM-like Higgs observed at 
the LHC and future prospects to discover other scalars.

Enlarging the scalar sector of the SM is a common practice in many theories (as, e.g., in Supersymmetry or 2 Higgs doublet models), it is 
thus no surprise that studies, either on the phenomenology or the experimental side, have already explored this possibility. For 
instance, LEP searches for a light SM Higgs provide us constraints on theories with an additional light scalar, while the LHC provides 
at present analyses for heavy Higgses. However interpreting those limits in a common framework with the measurements of the SM-like 
Higgs couplings has not yet become a standard, therefore the potential power of such studies can be improved if this possibility is 
considered.

The aim of the article is the following: we will introduce a parametrisation of a two scalars model, one of them corresponding to the 
observed Higgs state and one to a new scalar, and advocate for its use to facilitate the interface between theories and experimental 
results. We shall show in some specific theories how much the information we have on the SM-like Higgs affects the 
predicted behaviour of the other scalar. We consider the case of an additional scalar lighter than the SM-like Higgs: this choice is 
motivated by the fact that such searches are currently not handled by the LHC analyses, making it an interesting playground for testing 
possible scalars that would have escaped the LEP searches. In the rest of the paper we will refer to the scalar state corresponding to 
the observed resonance as ``the SM-like Higgs'', and the other one as ``the light scalar''.

The article is organised as follows: section \ref{sec:param} defines our parametrisation, section \ref{sec:models} describes the two specific models that we have 
chosen to exemplify the parametrisation, section \ref{sec:constraints} deals with the various constraints that we have imposed on such models and 
section \ref{sec:results} comments the different features of a possible light scalar search at LHC in both models. 

\section{Parametrisation of a light scalar} \label{sec:param}

The basic way to test a model is to compare the predicted rate of a process to an 
observed rate. However, matching the two is often difficult due to experiment-specific cuts and statistical combinations. By employing a 
generic parametrisation, this work needs to be done only once and the results are then available for any BSM model belonging 
to the class described by the parametrisation: in our case models with the presence of an extra light scalar and all other BSM states 
decoupled, at least within a certain approximation.

As we are interested in a lighter additional scalar and the Higgs boson is the only scalar particle in the SM, we will compare the 
effects of this new particle to those of a SM Higgs boson of the same mass. Therefore the parameters we use for both scalars are 
defined in a similar way: we normalise the couplings of each one to those of a SM Higgs with the same mass. This choice is not 
specially motivated by theoretical considerations, but is a simple way of choosing a common normalisation, which does not 
affect the generality of the approach.

Our main assumptions are that both scalars are $\mathcal{CP}$-even, custodial symmetric, according to which the couplings of the W and Z bosons to the two scalars scale with a common factor and that their couplings are universal in flavour:
all up-type or down-type quarks or leptons coupling factors. Finally, for particles that do not couple at tree-level to the Higgs, such as photons and gluons, we use a loop-inspired parametrisation.

\subsection{Tree-level modifiers}

The first kind of particles related to the Higgs Physics are obviously those which interact at tree-level with the Higgs boson in the 
SM, i.e. quarks, leptons and weak bosons. These interactions appear in the Lagrangian from the covariant derivative in the 
case of W and Z and from the Yukawa terms for fermions. 
In BSM models these interactions may differ from those of a SM Higgs boson although since we consider here only models with the same 
Lorentz structure as in the Standard Model, we can parametrise the difference by a ratio of couplings. As the interactions 
of both scalars will be altered, we define two sets of parameters as follows:
\begin{equation}
 \begin{aligned}
  g_{h_ibb}&=\kappa_{b,i}\ g^{SM}_{hbb}\hspace*{2cm} &g_{h_itt}&=\kappa_{t,i}\ g^{SM}_{htt}\\
  g_{h_i\tau^+\tau^-}&=\kappa_{l,i}\ g^{SM}_{h\tau^+\tau^-} &g_{h_iZZ,WW}&=\kappa_{V,i}\ g^{SM}_{hZZ,WW}
 \end{aligned}
\end{equation}
where $i = 1,2$ labels the two Higgs bosons $h_{1,2}$ ($h_2$ being the 126 GeV Higgs), and custodial symmetry implies that the couplings to the $W$ and $Z$ bosons are multiplied by the same modifiers $\kappa_{V,i}$.
Following our assumption of universality, the couplings of the lighter generations are scaled by the same modifiers as top, bottom and tau.

\subsection{Loop induced modifiers}

Among the main decay modes of the Higgs boson, two stand apart since they do not occur at tree-level: 
photons and gluons. As advocated in \cite{flament_1210}, we do not define the parameters measuring the effect 
of New Physics by the ratios of the partial widths (as done in \cite{YR3}) but rather by the ratios of the loop amplitudes. 
In this way, we can define parameters that only include the loop effects of new physics states, thus explicitly disentangling loop effects from the modification of tree level couplings.
For instance, the coupling to photons is affected by the $\kappa_{\gamma \gamma,i}$ parameter, defined as follows:
\begin{equation}
 \begin{aligned}
  \Gamma_{h_i\rightarrow \gamma\gamma} &\propto |\A_{W^{\pm}} + \A_t + \A_b + \A_{NP}|^2\\
  &\propto |\kappa_{V,i} \A_{W^{\pm}}^{SM}+ \kappa_{b,i} \A_b^{SM} +(\kappa_{t,i} + \kappa_{\gamma\gamma,i})\A_t^{SM}|^2,
 \end{aligned}
\label{eq:width}
\end{equation}
where $\A_{X}^{SM}$ is the amplitude of the loop of the particle X calculated with SM couplings for a Higgs boson of the mass of 
the scalar $h_i$. We took the top amplitude $\A_t$ as a reference amplitude for New Physics since it allows for an easy interpretation 
in terms of top partners, but it should be mentioned that this choice is purely arbitrary. Although the ratio of partial widths may seem 
more convenient on the experimental side, we choose this parametrisation to keep the loop structure visible and avoid correlations 
between $\kappa_{\gamma\gamma,i}$ and tree-level parameters ($\kappa_{V,i},\kappa_{t,i},\dots$) that would appear otherwise. We 
define similarly $\kappa_{gg,i}$ for the coupling to gluons. For more details see \cite{cacciapaglia_0901,flament_1210}.

\section{Examples of specific BSM models}\label{sec:models}

In order to show how our parametrisation applies to different models, 
we will probe two theories that have been the subject of many analyses 
dedicated to the Higgs phenomenology: the 2HDM (2 Higgs Doublet Model) which is an example of a typical description of an 
extended Higgs sector, and the NMSSM (Next to Minimal Supersymmetric Standard Model) which is a typical example of an ultra-violet 
completion of the low energy effective scalar sector. Due to their rich phenomenologies, a complete analysis is beyond the scope of this 
work: we only take them to illustrate how one can compare directly two widely different models through the parametrisation we 
suggest. We emphasise that the aim of our study is not to make statements on the phenomenology of 2HDM and NMSSM in general, 
but rather to show how the parametrisation can help in disentangling different specific models. Note that for both models we will 
impose $\ma>60$ GeV in order to avoid the decay $h_2\to h_1 h_1$. Furthermore, this mass range probably is the easiest 
region to search for a light scalar at the LHC, as signals at lower mass would be marred by very large backgrounds and would be severely affected by the triggers.

\subsection{Two Higgs Doublet Model}
A simple extension of the scalar sector of the SM is to add another doublet: this gives rise to the well known 2HDM, which has been 
thoroughly studied with respect to the LHC (with a lighter scalar in \cite{celis_1302,coleppa_1305,chang_1310}, and a heavier one in 
\cite{chiang_1303,barroso_1304,grinstein_1304,chen_1305,craig_1305,eberhardt_1305,belanger_1306,barger_1308,ko_1309,celis_1310,cheung_1310}). 
In the following, we will only highlight the details relevant to our study since the detailed equations can be found elsewhere 
(see \cite{branco_1106}, among others).

In particular the addition of a second doublet changes the physics of the whole scalar sector, particularly the potential which now takes the 
following form:
\begin{eqnarray}
\mathcal{V} & = & m_{11}^2 \Phi_1^{\dagger} \Phi_1 + m_{22}^2 \Phi_2^{\dagger} \Phi_2 - \left[m_{12}^2 \Phi_1^{\dagger} \Phi_2 
+ h.c.\right] \nonumber\\
& &+ \frac{1}{2} \lambda_1 \left(\Phi_1^{\dagger} \Phi_1\right)^2 + \frac{1}{2} \lambda_2 \left(\Phi_2^{\dagger} \Phi_2\right)^2 + 
\lambda_3 \left(\Phi_1^{\dagger} \Phi_1\right)\left(\Phi_2^{\dagger} \Phi_2\right)+ \lambda_4 \left(\Phi_1^{\dagger} \Phi_2\right)
\left(\Phi_2^{\dagger} \Phi_1\right) \nonumber \\
& &+ \left\{\frac{1}{2} \lambda_5 \left(\Phi_1^{\dagger} \Phi_2\right)^2 + 
\left[\lambda_6 \left( \Phi_1^{\dagger} \Phi_1 \right) + \lambda_7 \left( \Phi_2^{\dagger} \Phi_2 \right) \right] \left(\Phi_1^{\dagger} 
\Phi_2\right) + h.c. \right\}\,.
\end{eqnarray}
As the potential is real, $m_{11}^2,\ m_{22}^2,\ \lambda_{1-4}$ need to be real, while $m_{12}^2,\ \lambda_{5-7}$ may be complex. As we limited ourselves to a $\mathcal{CP}$-conserving Higgs sector, we will here consider all parameters to be real.
Furthermore, by imposing a $\mathbb{Z}_2$ symmetry, we can set $\lambda_{6,7} = 0$ and $m_{12}^2 = 0$. 

Since the model now has two scalar doublets $\Phi_1$ and $\Phi_2$, the EWSB is not necessarily achieved by having only one of the two doublets acquiring a 
$vev$, but a combination of the two. We can identify this combination by rotating to a basis in which only one of the doublets acquires a 
$vev$. This rotation is parametrised by an angle known as $\beta$. After symmetry breaking, instead of the one Higgs boson in the 
standard model, we get five scalars from the remaining degrees of freedom of the doublets: two $\mathcal{CP}$-even neutral states $h_{1,2}$, 
one $\mathcal{CP}$-odd neutral state $A^0$ and two charged ones $ H^{\pm}$, charge conjugates. As for the quadratic part of the 
Lagrangian, the $\mathcal{CP}$-odd and charged Higgses are necessarily mass eigenstates, but there is still possible 
mixing for the two $\mathcal{CP}$-even scalars. As we rotate the basis once again from the $vev$-acquiring-basis to the mass basis, 
we introduce another rotation angle, $\talpha$. This angle is related to the commonly defined angle $\alpha$ by 
$\talpha = \beta-\alpha$. All parameters from the potential can then be translated into the parameters in that last basis: 
\begin{gather}
  \lambda_1,\ \lambda_2,\ \lambda_3,\ \lambda_4,\ \lambda_5,\ \lambda_6,\ \lambda_7,\ m_{11}^2,\ m_{22}^2,\ m_{12}^2 \nonumber\\
  \Updownarrow\\
  m_{h_1},\ m_{h_2},\ m_{A^0},\ m_{H^+}=m_{H^-},\ \tb, \sin(\beta - \alpha) = \sa,\ v,\ \lambda_6,\ \lambda_7,\ m_{12}^2\nonumber
\end{gather}
In the following we will probe the parameter space of the model by a numerical scan over the parameters.
We shall probe the following ranges:
\begin{center}
\begin{tabular}{c|c|c|c|c|c|c}
$m_{h_1}$ (GeV)& $m_{h_2}$(GeV) & $m_{A^0}$(GeV) & $m_{H^+}$(GeV) & $\sa$ & $\tb$ & $v$(GeV) \\\hline
$[70,120]$ & $ 125 $ & $ [300,1000]$ & $ [300,1000] $ & $ [-1,1] $ & $[0.35,50]$ & $246$ 
\end{tabular}
\end{center}
while $\lambda_6$, $\lambda_7$ and $m_{12}^2$ are zero as we stick to the pure $\mathbb{Z}_2$ symmetric case.

The Yukawa sector of the SM is also affected, as each Yukawa coupling can be replaced by two couplings involving the two doublets. In order to avoid tree-level FCNCs, all particles of the same kind should couple to the same doublet (for instance, all up-type quarks couple to $\Phi_1$), 
thus we are left with 4 types of 2HDM models, distinguished by which doublet the down-type quarks and charged leptons couple to, as summarised in the following table:
\begin{center}
\begin{tabular}{c|c|c|c|c}
 & \multicolumn{4}{c}{Type} \\\cline{2-5}
 & I & II & III & IV \\\hline
Up Quarks & $ 1 $ & $ 1 $ & $ 1 $ & $ 1 $\\\hline
Down Quarks & $ 1 $ & $ 2 $ & $ 2 $ & $ 1 $\\\hline
Leptons & $ 1 $ & $ 2 $ & $ 1 $ & $ 2 $
\end{tabular}
\end{center}
In particular, we have in all types for up-type quarks the following couplings to the two $\mathcal{CP}$-even Higgses:
\begin{equation}
 \begin{aligned}
  \kva & = \sa & \kvb & = \ca \\
  \kta &= \sa + \frac{\ca}{\tb}= \kva + \frac{\kvb}{\tb} \qquad &\ktb &= \ca - \frac{\sa}{\tb} = \kvb - \frac{\kva}{\tb},
\label{eq:2hdm}
 \end{aligned}
\end{equation}
where we also show the relation in terms of $\kappa$ parameters. Couplings to down-type quarks (leptons) will be the same in 
type I and IV (I and III) and be obtained by exchanging $\tb\leftrightarrow -1/\tb$ in type II and III (II and IV). Another useful relationship to understand correlations 
between parameters is that, regardless of the type, we have for any fermion family:
\begin{equation}
 \kfa\kva+\kfb\kvb=1\,.
  \label{eq:kvkf}
\end{equation}

\subsection{NMSSM}
The NMSSM is the simplest extension of the MSSM (Minimal Supersymmetric Standard Model, which is itself the minimal extension of 
the Standard Model that allows for Supersymmetry), obtained by adding a singlet chiral superfield to the matter spectrum. There are 
several reasons to prefer the NMSSM as compared to the MSSM: some theoretical (NMSSM solves what is known as the $\mu$ 
problem), and some experimental since its phenomenology is more flexible. In particular, our study deals with the case 
where the second CP-even Higgs is the SM-like Higgs, and this possibility is now extremely constrained if not excluded in the MSSM by experimental 
data (see e.g. \cite{arbey_1211}). The implication of Higgs measurements has been discussed in many 
publications (see \cite{king_1211,cheng_1304,barbieri_1304,beskidt_1308,choi_1308}, among others), in particular with a 
light scalar in\cite{belanger_1210,cerdeno_1301,christensen_1303,barbieri_1307,cao_1309,badziak_1310}. We choose here to deal with a NMSSM with a 
$\mathbb{Z}_3$ symmetric superpotential. We will not detail its construction, the interested reader can find it in one of the many 
reviews on the subject (for instance \cite{ellwanger_0910}). We have used \texttt{NMSSMTools-4.0} \cite{nmssmtools} to compute the 
relevant observables. Note that we do not take directly the Higgs signal strengths, but instead we use the reduced couplings, and 
recompute signal strengths ourselves.

We shall use the following parameter space: we choose common masses for sfermions: $M_{\tilde{f}}=500$ GeV for sleptons and
$M_{\tilde{q}}=2$ TeV for squarks, together with vanishing trilinear couplings, except for $A_t$. This ensures that we are above LHC limits and at 
the same time decouple them from Higgs physics (apart for the corrections to the Higgs masses). We choose gaugino masses as $M_1=100$ GeV, 
$M_2=500$ GeV and $M_3=1$ TeV: since we are not interested in Dark Matter observables, we can leave those fixed. The reason why 
we free ourselves from this constraint is that a correct value of the relic density can often be achieved without changing the Higgs 
Physics, for instance simply by adjusting $m_{\tilde{\chi}_1^0}$ (which can be done by playing on $M_1$) so that we sit on a resonance 
($Z$, $h_1$ or $h_2$). In particular we have checked that even with $M_1=100$ GeV, many of our points exhibit a mostly bino 
(i.e. with a bino fraction larger than 80\%) and partly singlino and Higgsino neutralino -- which can hence benefit from such a 
resonance -- and that selecting those points does not affect our conclusions on the Higgs sector. We leave thus a refined analysis to a future study. 
The remaining parameters will be varied in the following range:
\begin{center}
\begin{tabular}{c|c|c|c|c|c|c}
$\tb$ & $\mueff$ & $\lambda$ & $\kappa$ & $A_\lambda$ & $A_\kappa$ & $A_t$ \\\hline
$[1,50]$ & $[100,600]$ (GeV) & $[0,0.75]$ & $[0,0.3]$ & $[-1,1]$ (TeV) & $[-1,1]$ (TeV) & $[-4,4]$ (TeV)
\end{tabular}
\end{center}
We keep $A_t$ as a free parameter as it drives most of the radiative corrections to the Higgs masses. We will impose the constraint 
$\mb=125.5\pm3$ GeV, where the uncertainty is purely theoretical. A flat scan of such a parameter space would give very few points 
abiding by these constraints (most of the time $h_2$ would be heavier) so we have used a genetic algorithm based scan: the basic 
idea is to start from a random population of points in the parameter space that we subsequently evolve by mutations and 
crossovers and finally select on their ability to pass the constraints that we impose on the phenomenology (in particular 
$\mb=125.5\pm3$), and we repeat this operation until we find enough correct points.

As an aside, note that we do not impose constraints on the $\mathcal{CP}$-odd scalars $a_1,\ a_2$ and on the heaviest $\mathcal{CP}$-even $h_3$ from direct searches, choice that we justify by the fact that 
$a2$ and $h3$ are heavy (above 1 TeV) and that they take most of the $\tb$ enhancement effect, so that prospects for discovering $a_1$ 
(which is lighter) are rather dim. We also stick to the case $\maa>635$ GeV, to prevent decays of $h_1,h_2$ to $a_1$.

\section{Constraints}\label{sec:constraints}
In order to have realistic points for our two specific models (2HDM and NMSSM), we have applied constraints of two 
kinds: model dependent ones and model independent ones. The latter are purely related to the neutral Higgs phenomenology 
and can be dealt with at the level of the parametrisation, while the former stems from other sectors and will thus be different in each 
model. In the 2HDM case, we will check for perturbativity of the couplings and stability of the potential on the theory side, and on the 
experimental side the electroweak precision test (S,T and U parameters), the muon anomalous magnetic moment ($\Delta a_\mu$) and 
the following observables from flavour physics: $\bsg$, $\bsmumu$, $\dmd$. Those quantities have been evaluated using \thdmc\ 
\cite{2hdmc} and \superiso\ \cite{superiso}, and the experimental bounds are given in table \ref{tab:constraints}. For the NMSSM, we 
have also used electroweak precision tests, muon anomalous moment and flavour physics with $\bsg$, $\bsmumu$. Note that in both 
cases we do not include direct search for charged Higgs since it is too heavy to be constrained ($\mhp>300$ GeV).
\begin{table}[!h]
\begin{center}
 \begin{tabular}{c|c}
  Observable & Exp. bound\\\hline
  \{S,T,U\} & \{$[-0.10,0.11]$,$[-0.10,0.13]$,$[-0.03,0.19]$\} \cite{Beringer_2012_ew_const}\\
  $\Delta a_\mu$ & $\Delta a_\mu<4.5\times 10^{-9}$ \cite{bennet_2006}\\
  $\bsg$ & $\text{Br }\bsg=3.43\pm 0.9 \times 10^{-4}$ \cite{hfag}\footnotemark \\
  $\bsmumu$ & $1.5\times 10^{-9}<\text{Br }\bsmumu<3.3\times 10^{-9}$ \cite{bsmumu}\\
  $\dmd$ & $\dmd=0.53\pm 0.16 \text{ps}^{-1}$ \cite{hfag}\footnotemark[\value{footnote}]
 \end{tabular}
\caption{\label{tab:constraints}{\em Experimental constraints to be applied on the two specific models.}}
\end{center}
\end{table}
\footnotetext{We have incorporated the theoretical uncertainty obtained in \cite{mahmoudi_0907,misiak_2007} to the experimental 
measurement, treating it as a nuisance parameter.}

Turning to the constraints on the neutral Higgs sector, our main interest will be in the LEP limits and the LHC measurements. It turns out 
that the relevant observables can be completely determined by the value of the $\kappa_{X,i}$ parameters, plus the mass of the light 
scalar $h_1$, which has the important consequence that the test can be carried out at the level of the parametrisation, without 
reference to the actual model. Concerning the LEP limits, we used the program \hb\ \cite{higgsbounds}\footnote{Note that the current version (4.1.0) of \hb\ does not include the most sensitive limit on $h\to\text{hadrons}$, hence we included ourselves the results from \cite{lep2j}}, which takes as input the 
effective couplings. Those are either directly related to $\kappa_{X,1}$ parameters for tree-level couplings, or through the formulae in Eq. 
\ref{eq:width} for loop-level couplings. Note that we used the simplifying assumption that the ratio of gluon-fusion production cross-sections is equal to the ratio of the partial widths into gluons since it saves us the integration over parton distribution functions, and that 
the approximation holds up to 10\%. For the couplings of $h_2$ measured at the LHC, we build a $\Delta\chi^2(\kappa_{X,2})$ test 
based on the procedure introduced in \cite{flament_1210}: that is to say, for each decay mode 
($XX\in\{ZZ,WW,\gamma\gamma,\bar bb,\bar\tau\tau\}$) we take the 2D contours in the $(\mu_{ggh},\mu_{VBF})$ plane, as given by the ATLAS 
\cite{atlas_couplings} and CMS \cite{cms_couplings} collaborations, then compute the $\Delta\chi^2_{XX}$ associated to any given 
point and finally sum over final states. We define allowed points as points in the 95\% confidence level volume.

\section{Results}\label{sec:results}
We shall now apply our parametrisation to the following specific question: given what we know about the SM-like Higgs and the null 
result for searches at LEP, what are our best chances to find the light scalar $h_1$ at the LHC? The answer depends on the 
hypothesised model, and thus we will compare 2HDM and NMSSM. Note that if we do not specify a model the question has no 
practical interest: indeed if $h_1$ and $h_2$ are uncorrelated, what we know on $h_2$ couplings has no influence on those of $h_1$, 
and since LEP and LHC use different production mechanisms (relying on the coupling to electroweak boson for LEP and 
gluons for LHC\footnote{LHC is also sensitive to VH or VBF production, but one usually 
expect gluon fusion to be dominant.}) the null searches at LEP impact very mildly LHC prospects.

On the experimental side, it is important to note that nearly all LHC analyses start at $m_H>110$ GeV (at least we are not aware of any 
public results lowering this bound). While this approach is valid when looking for the SM Higgs since LEP already ruled it out for lower 
masses, it has no support when looking for a non-SM Higgs such as the one that we are dealing with. While such a search certainly 
presents new challenges from the experimental point of view (not all final states may be exploitable), we do think that it is worth 
studying, since many models still allow for light additional scalars. After all, if experimental collaborations dedicate full analyses to the 
high-mass Higgs search, they could also bring up the low-mass Higgs search. We will not dwell more on the subject of experimental 
sensitivity, and, for the sake of concreteness, we shall focus on the search channel $h_1\to\gamma\gamma$.

\begin{figure}[!t]
\begin{center}
\begin{tabular}{ccc}
\includegraphics[width=0.3\textwidth]{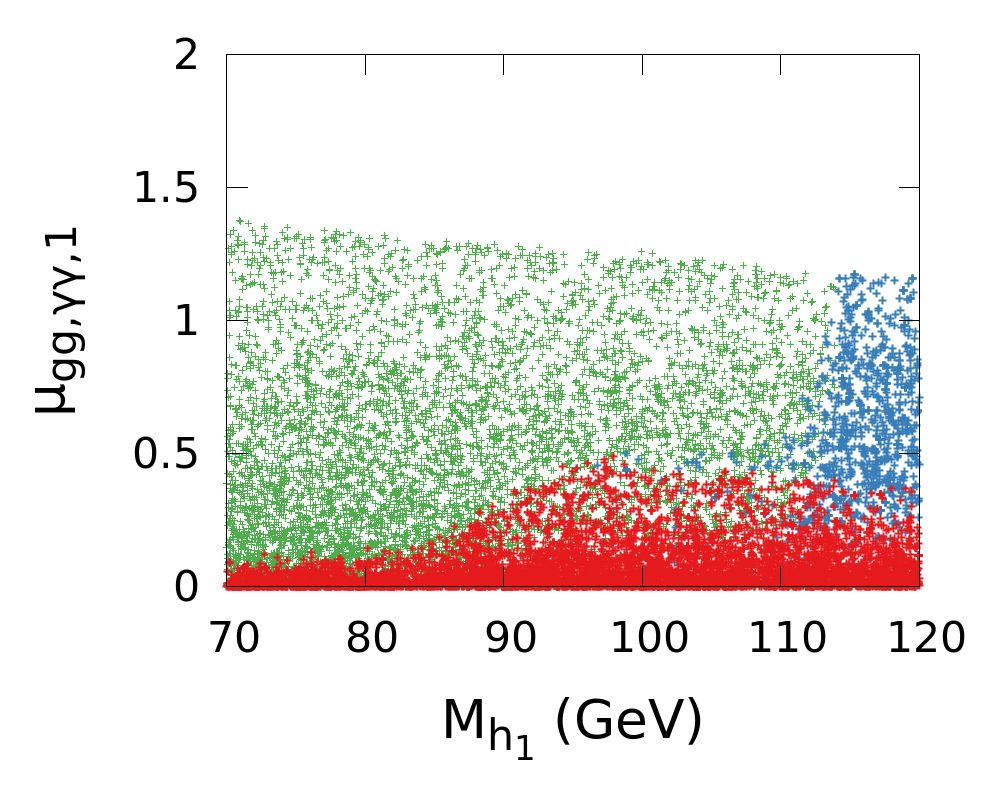}&
\includegraphics[width=0.3\textwidth]{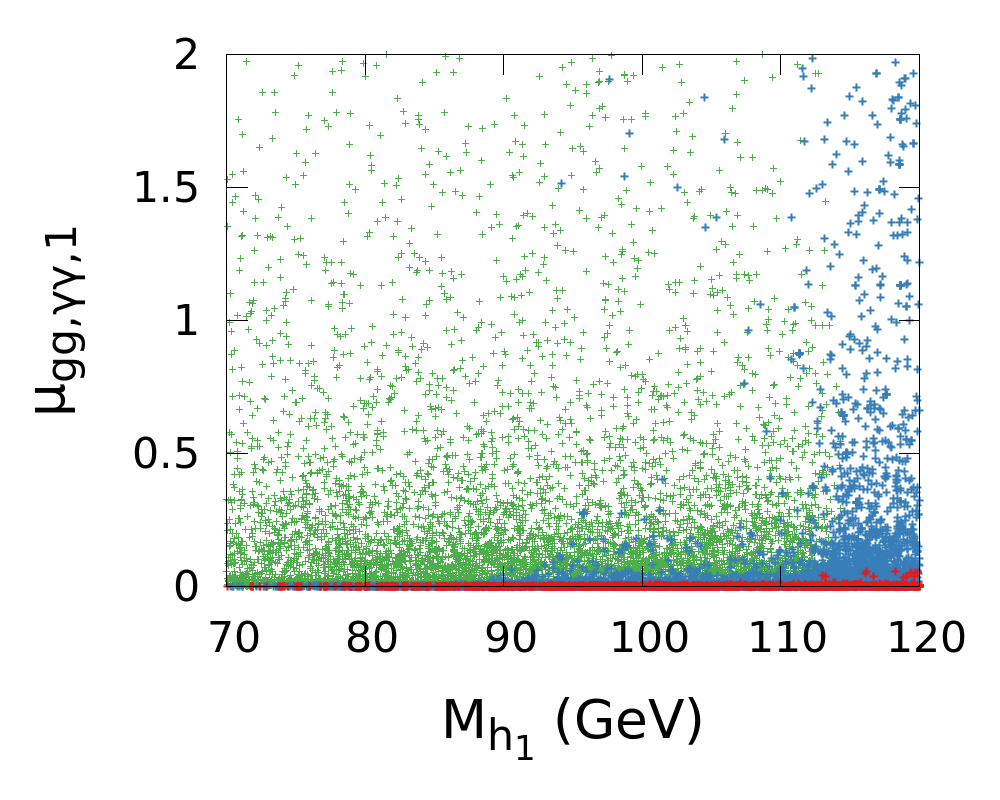}&
\includegraphics[width=0.3\textwidth]{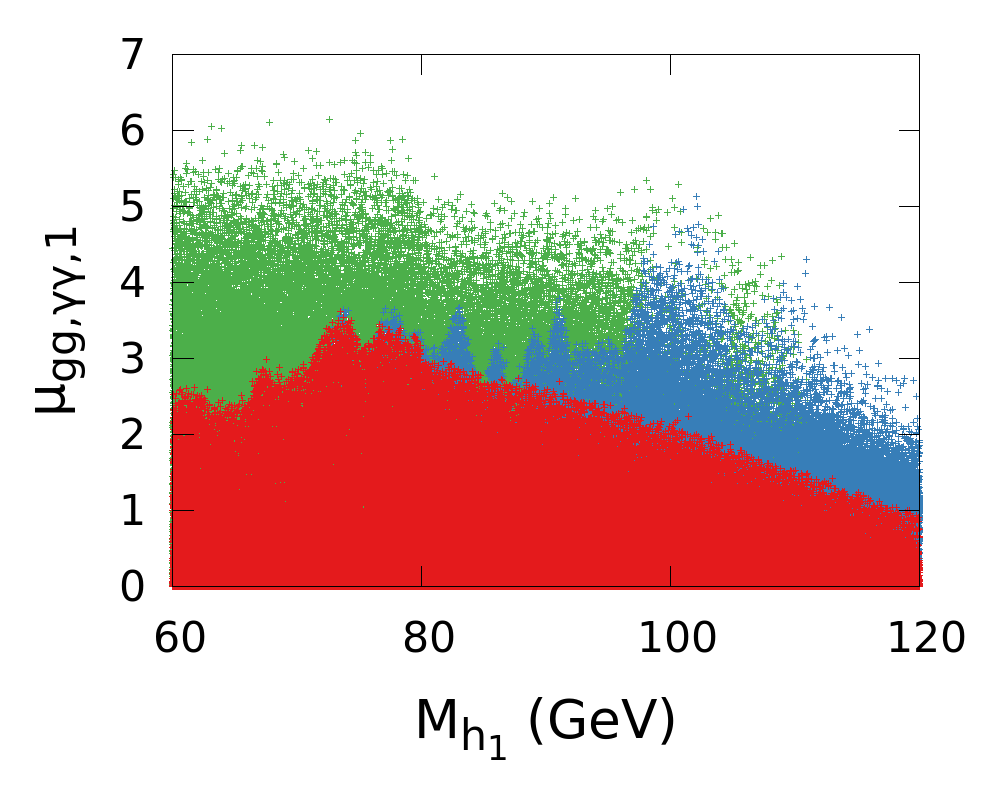}\\
\includegraphics[width=0.3\textwidth]{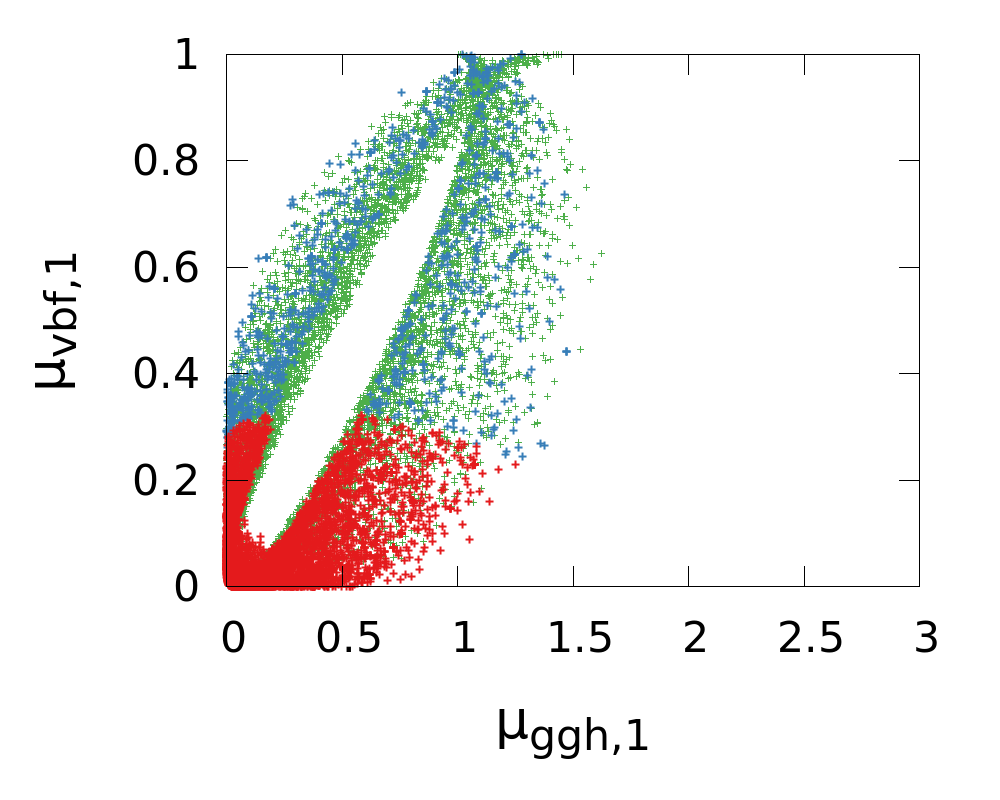}&
\includegraphics[width=0.3\textwidth]{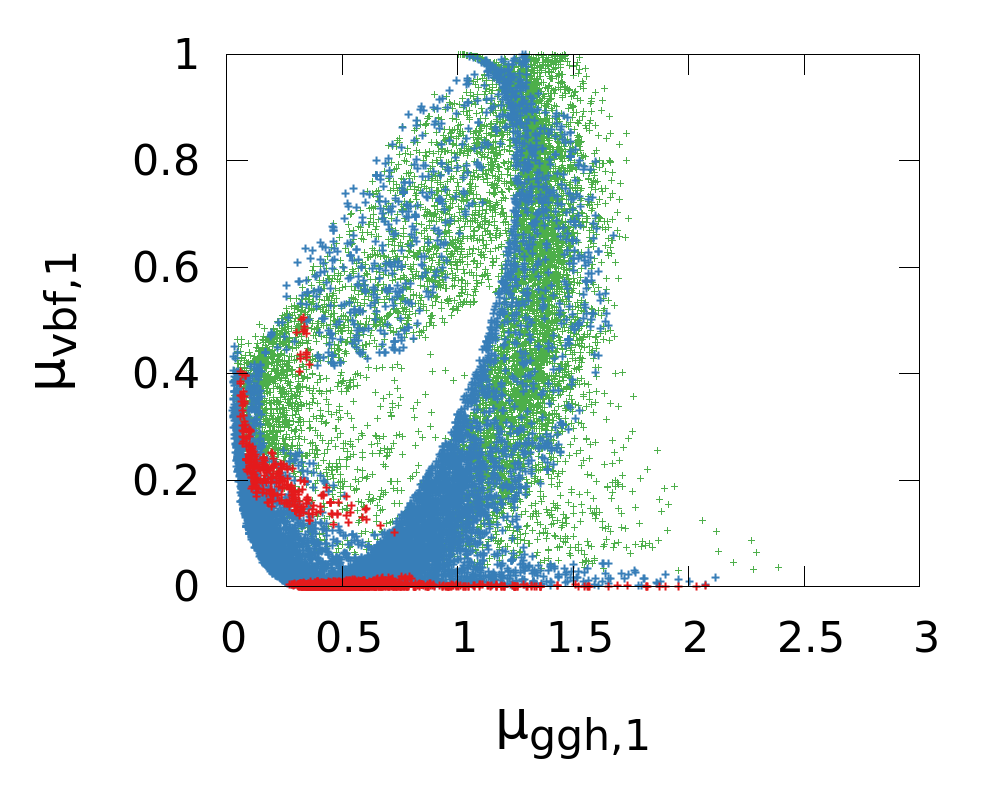}&
\includegraphics[width=0.3\textwidth]{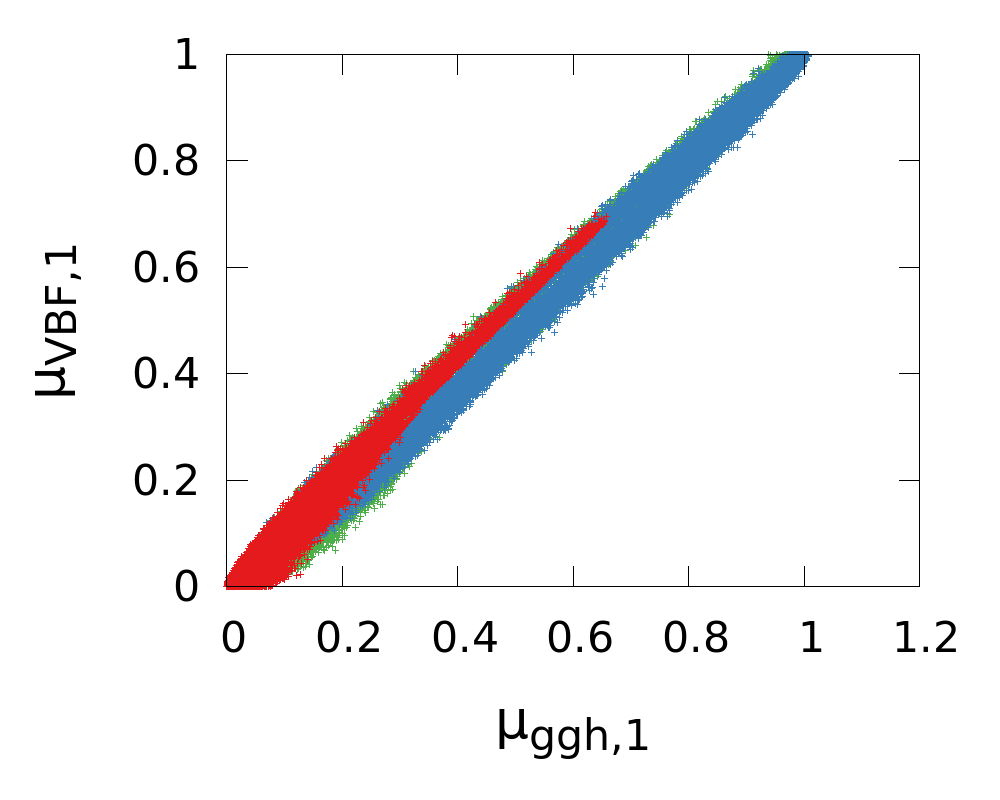}\\
2HDM(I)&2HDM(II)&NMSSM
\end{tabular}
\end{center}
\caption{{\small Top: signal strength in the $gg\to h_1\to\gamma\gamma$ channel. Bottom: $\ggh$ production mode versus $\VBF$, 
both normalised to the SM. The colour code is the following: Green (light grey) points are all points passing flavour and theoretical 
constraints, blue points (grey) are a subset of those which also pass LEP constraints on $h_1$ and red (dark grey) points pass in 
addition the LHC couplings constraint on $h_2$.}}
\label{fig:mugam}
\end{figure}

We show in figure \ref{fig:mugam} the reach of the signal strength $\mu$ in the channel $gg\to h_1\to\gamma\gamma$ for both 
models. This is an important piece of information, since it already tells us that the LHC has the power to discriminate some of the scenarios: 
indeed some points of the NMSSM and, to a lesser extent, of type-I 2HDM can reach high values ($\mu_{gg\to h_1\to\gamma\gamma}$ 
can reach 3 in the NMSSM and 0.5 in 2HDM-I), whereas 2HDM-II shows much more modest values. We remind that the 95\% excluded signal 
strength reported by ATLAS and CMS lies around $1$ at $m_H=110$ GeV, so points with $\ma>100$ and $\mu_{gg\to h_1\to\gamma
\gamma}>1$ are probably probed by the current data-set. The colour code is the following: green points pass all ``non-Higgs'' 
constraints (in particular, flavour observables), blue points also pass the LEP constraints and finally red points are, in addition to that, 
compatible at 95\% with the SM-like Higgs couplings measured at the LHC.
It is also worthwhile to compare the production cross-section through gluon fusion and through Vector Boson Fusion (VBF) or
 associated vector boson production\footnote{Since our parametrisation respects custodial symmetry, we have $\mu_\VBF=\mu_\VH$.} 
(VH), as shown in the plots on the second row of figure \ref{fig:mugam}: as the two have different kinematics properties, the sensitivity towards a VBF dominated sample may 
differ from the one of a gluon-fusion dominated one. We see that in the 2HDM case, we can have very different composition whereas in 
the NMSSM the composition will be the one of the SM.

\begin{figure}[!t]
\begin{center}
\begin{tabular}{ccc}
\includegraphics[width=0.3\textwidth]{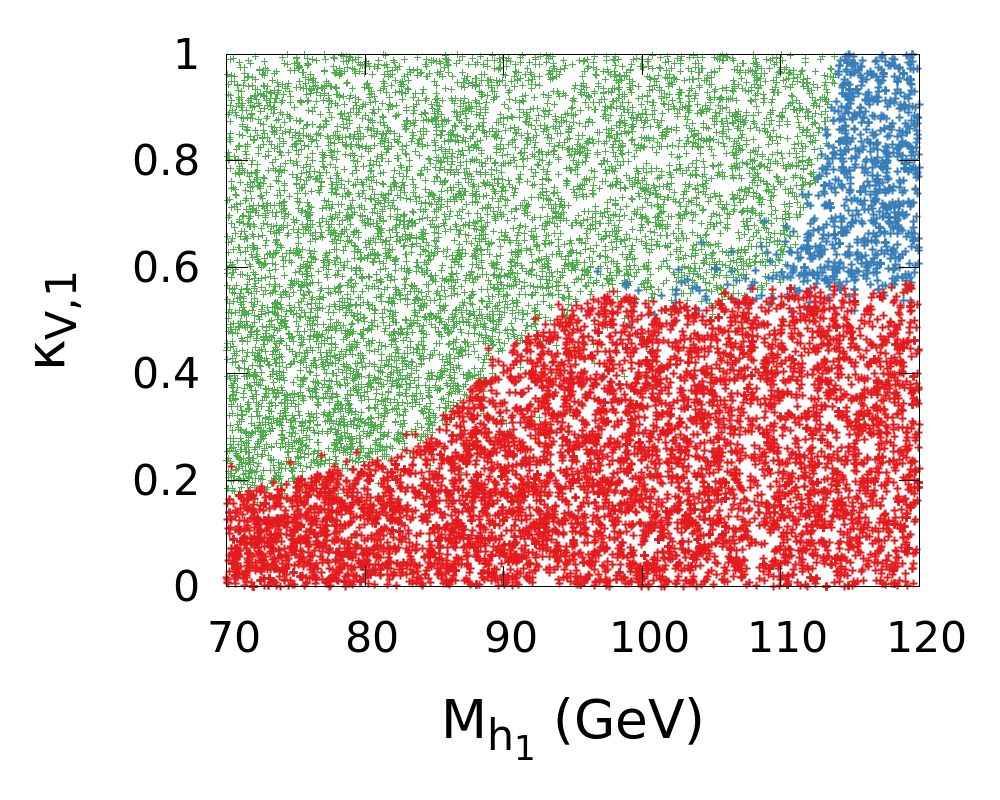}&
\includegraphics[width=0.3\textwidth]{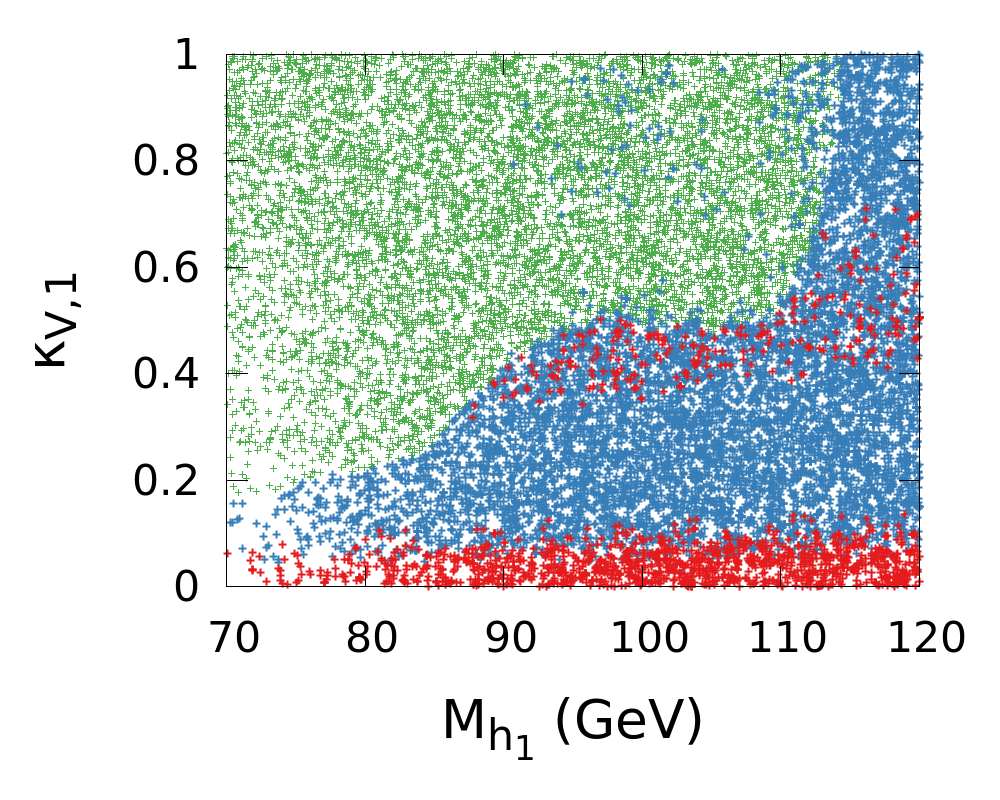}&
\includegraphics[width=0.3\textwidth]{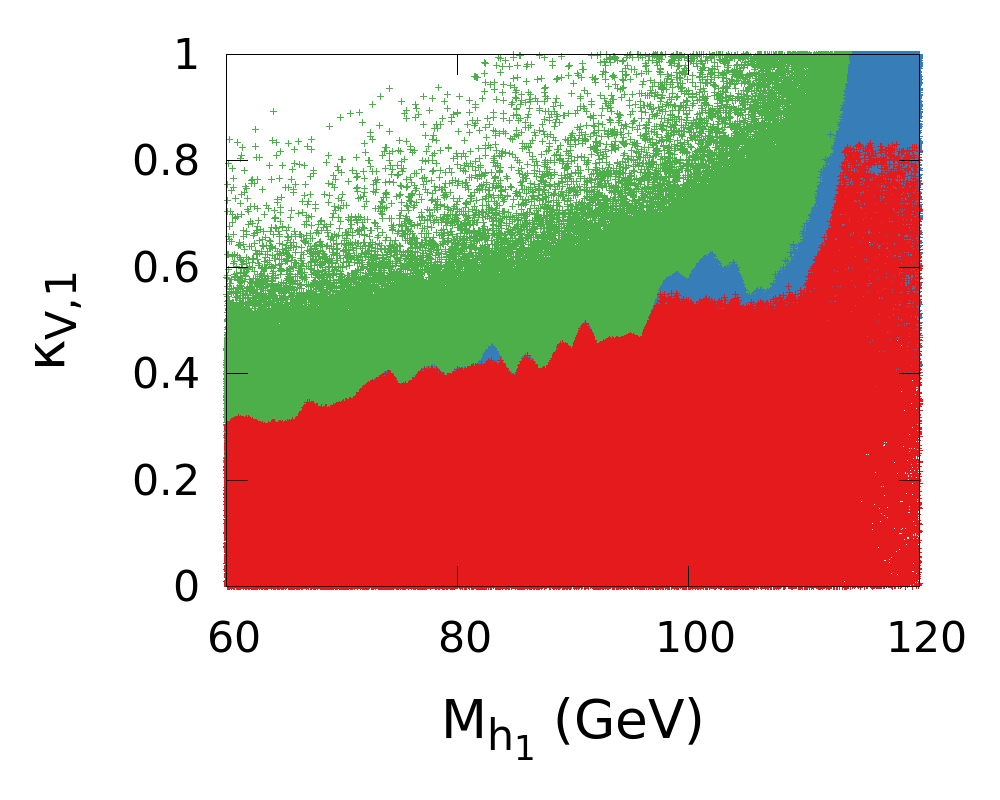}\\
\includegraphics[width=0.3\textwidth]{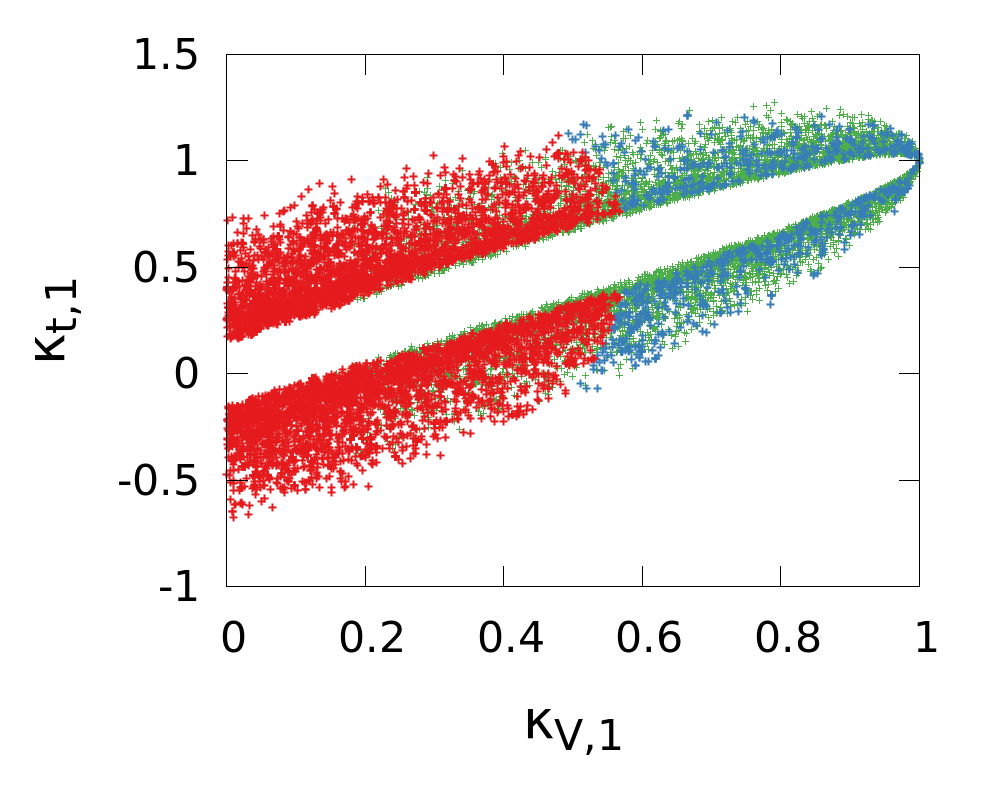}&
\includegraphics[width=0.3\textwidth]{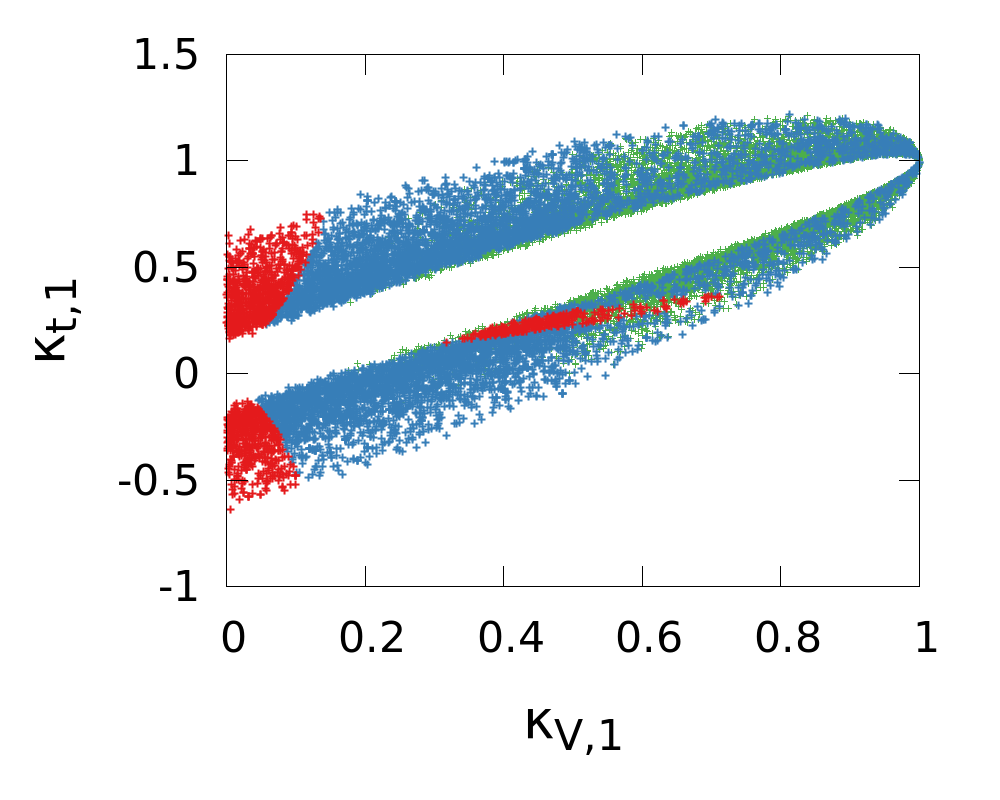}&
\includegraphics[width=0.3\textwidth]{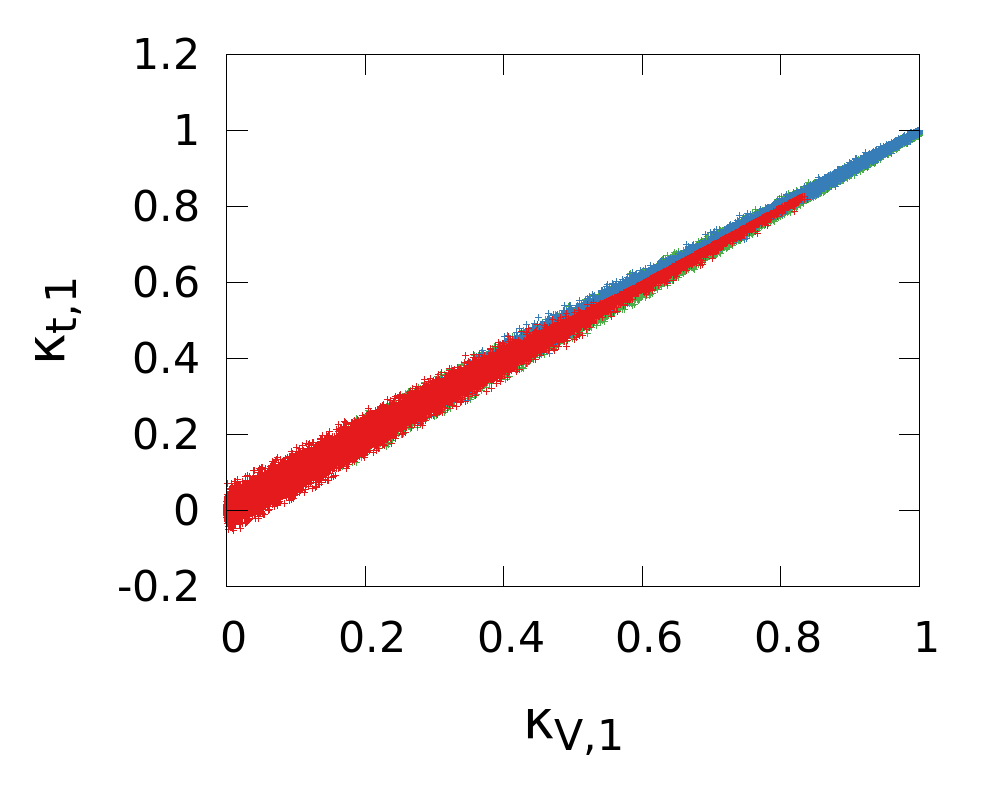}\\
\includegraphics[width=0.3\textwidth]{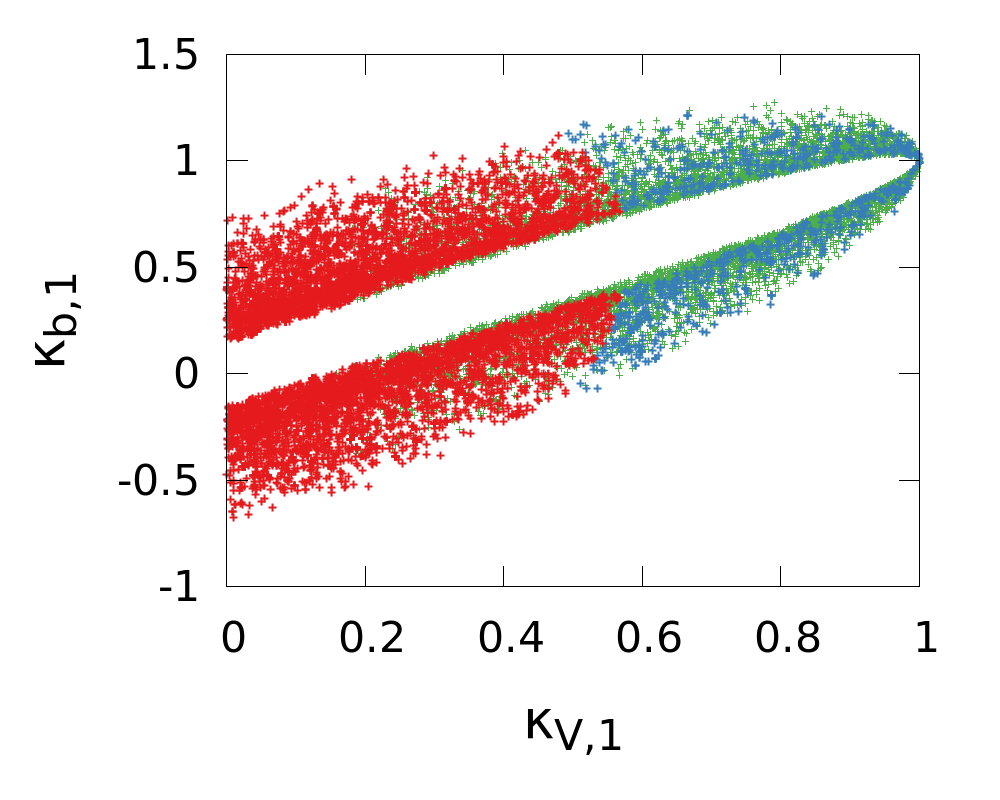}&
\includegraphics[width=0.3\textwidth]{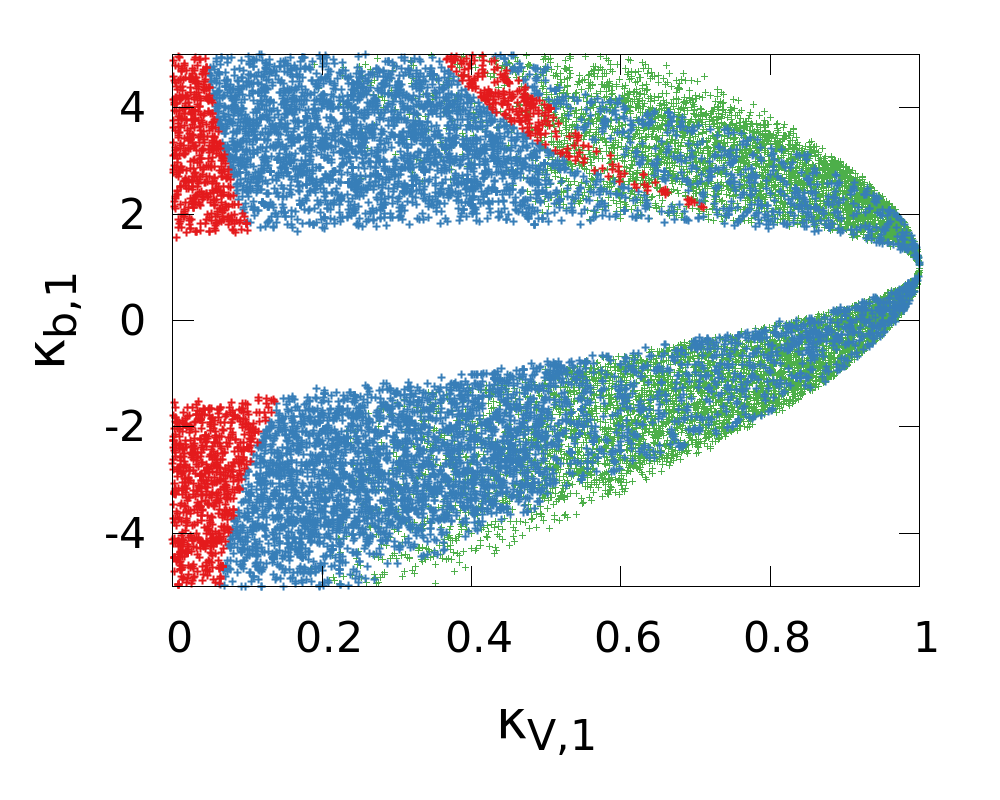}&
\includegraphics[width=0.3\textwidth]{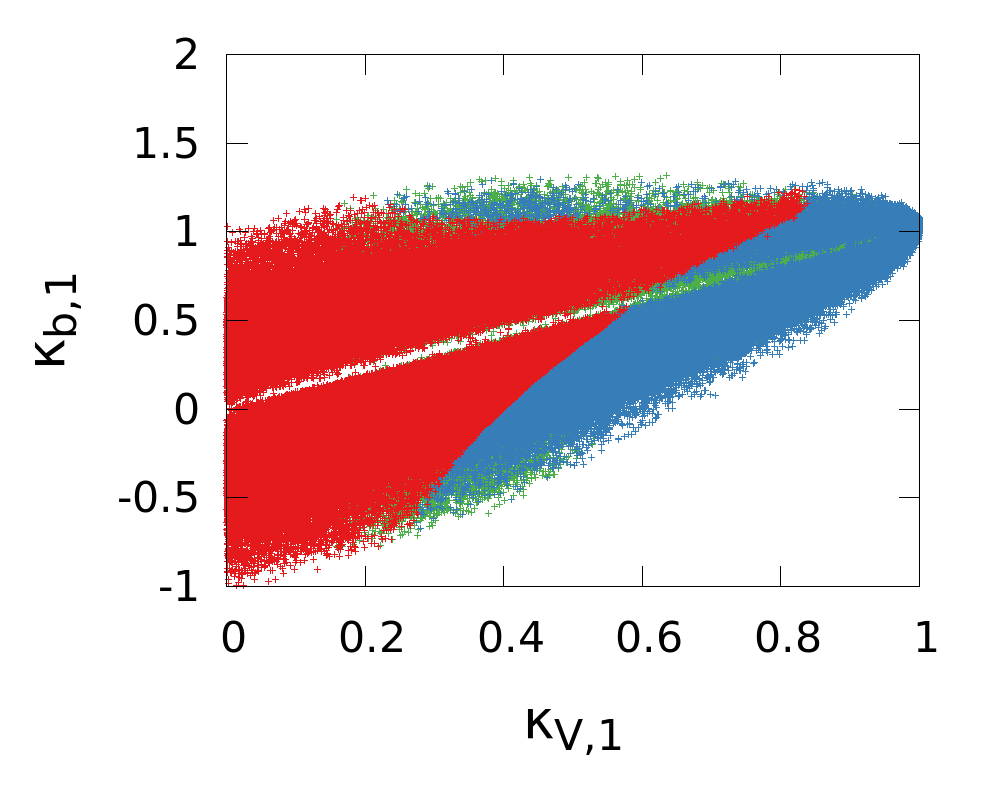}
\end{tabular}
\end{center}
\caption{{\small Correlations between $\kva,\kta,\kba$ and $\ma$: left column corresponds to 2HDM(I), middle one to 2HDM(II) and 
right one to the NMSSM. Colour code identical to figure \ref{fig:mugam}}}
\label{fig:kappas}
\end{figure}

The various features that appear at the level of signal strengths are better understood at the level of the $\kappa_{X,i}$ parameters. 
Thus, we show on the top row plots of figure \ref{fig:kappas} the relation between $\ma$ and $\kva$, making the LEP exclusion explicit. 
Since LEP is sensitive to all possible final states of $h_1$, its exclusion directly scale with the production through associated vector 
boson. It turns out that the NMSSM can weaken the bounds at low $\ma$: this stems from the fact that we can have a suppressed 
couplings to down-type quarks without a suppressed couplings to up-type quark (a situation that cannot be reproduced in 2HDM-I and 
that is forbidden by flavour tests in 2HDM-II), leading thus to an enhancement of the branching fraction to jets where the limit is 
weaker than for $\bar bb$ final state. We note that the LHC constraint also impacts this plot: this is due to the relation $\kva^2+\kvb^2=1$ 
which is exact in 2HDM, and turns out to hold to a very good approximation in the NMSSM\footnote{For all our points, $h_3$ turned out 
to be very heavy and to play no role in EWSB.}. As a short parenthesis, let us notice that in the case of 2HDM-II, some values of $\kva$ 
are forbidden: this is due to an interplay between flavour and LHC constraints. First, let us point out that the main contribution to the 
flavour observables in the 2HDM comes from the charged Higgs: it turns out that in our set-up, where both $h_1$ and $h_2$ are light, 
the charged Higgs cannot decouple if the perturbativity of the Higgs potential is to be preserved. Moreover, it occurs that since the 
coupling of the charged Higgs to up-type quark, which is independent of the Yukawa pattern, has a dominant impact on flavour 
observables, all types end up with a similar constraint on $\tb$, namely $\tb\in[2,12]$. Then, we deduce from equation \ref{eq:2hdm}, 
that in order to have $|\kbb|\approx 1$\footnote{The approximation is rather loose, \textit{i.e.} $|\kbb|= 1\pm 0.5$ is compatible with LHC 
Higgs couplings} in type II, we need $\kvb\approx 2/\tb$. Thus the constraint on $\tb$ from flavour physics translates to a constraint on 
$\kva$ due to the Higgs coupling data. Incidentally, note that the NMSSM avoids this constraint thanks to its third scalar: the latter, heavy, 
will take most of the $\tb$ enhancements, making thus the relation between $\tb$ and $\kba$ not as strict as in the 2HDM-II.\\

We also learn from figure \ref{fig:kappas} that the correlations between $\kva$ and $\kta$ differ significantly between the two 
models: in the NMSSM they have the same behaviour (which explains thus why $\mu_\VBF$ behaves the same way as $\mu_\ggh$) 
while in the 2HDM, $\kta$ can reach high values even for $\kva\approx0$.\\

A crucial point in disentangling the phenomenology of different models is through the coupling to $b$ quarks, which is parametrised by 
$\kba$: we show the $(\kva,\kba)$ plane in the bottom row of figure \ref{fig:kappas}. One notices that in the 2HDM-II case, there is no 
possibility for $\kba$ to vanish if $\kva$ does not, situation that differs significantly from 2HDM-I or the NMSSM. Such a feature would not come 
unnoticed at the LHC: indeed, since the width of a SM light Higgs is dominated by the $\bar bb$ decay, having $\kba\approx0$ will 
enhance all other decay modes, in particular the $\gamma\gamma$ one. In fact all points with a signal strength in the 
$gg\to h_1\to\gamma\gamma$ channel higher than 1 in figure \ref{fig:mugam} benefits from this mechanism. The largest enhancements also benefit from a non-vanishing $\kgama$ parameter: in particular some of the NMSSM points are noticeably enhanced by the chargino contribution. Thus in this case the significant difference of signal strengths from various theories can be explained in a 
universal way through the correlations of the $\kappa$ parameters, without recourse to the underlying UV completion.

Another feature of 2HDM-II shows up in the $(\kva,\kba)$ and $(\kva,\kta)$ planes of figure \ref{fig:kappas}: we see that the 
points allowed by LHC constraints for which $\kva$ does not vanish form lines which are either $\kta=\kva/2$ or $\kba=2/\kva$. This can be understood from eq.\ref{eq:kvkf}: approximating $|\kvb|\approx1-\kva^2/2$ and $|\kfb|\approx1$ (which is the limit of $h_2$ being completely SM-like with respect to 
fermions), one obtain the two solutions $\kfa=\kva/2$ or $\kfa=2/\kva$, depending on the sign of $\kvb\kfb$. For the time being, LHC 
constraints on $\kbb$ or $\ktb$ are not strong enough to justify the approximation $|\kfb|\approx1$, which is why there is no such lines 
in type I. However in type II, the Yukawa structure is such that, for $\tb>1$, small variations of $\ktb$ will correspond to large variations 
of $\kbb$. The later being not too loosely constrained\footnote{We recall that the main constraint on $\kbb$ does not come from the 
$VH\to V\bar bb$ analysis, but rather from the total width of the Higgs, which affect all signal strengths.}, the former will stay in a narrow 
band. This narrow band will thus translate to a line $\kta=\kva/2$, as foreseen by our earlier argument. Note that type III exhibits a very 
similar behaviour as of type II, while type IV is in-between type I and type II: in this case $\ktb$ and $\kbb$ vary the same way, but now 
$\klb$ exhibits large variations, and will thus constrain them. The LHC constraint on $\klb$ being looser than $\kbb$, the allowed bands 
will be broader.

\section{Conclusion}
The possibility of an extra scalar, lighter than the 126 GeV state already observed, is still compatible with current experimental data in 
different models of New Physics, and our study suggests that dedicated analysis of this mass range at the LHC are an interesting 
possibility for the experimental collaborations. If such an experimental search is carried out, it would be a crucial point to use a common 
parametrisation for the coupling of both scalars and to keep this parametrisation as general as possible. We have shown that the one 
we are suggesting (a straightforward extension of what is currently done for the 126 GeV state alone) could fulfil this goal, and comes 
with the advantage of being easily related to purely experimental quantities (such as cross-sections) and at the same time allowing a 
clear disentangling of some specific models in terms of correlations of the different $\kappa$ parameters. In addition to that, some 
effects at the experimental level, such as an enhancement in the channel $gg\to h_1\to\gamma\gamma$, could be explained directly in 
terms of the allowed values for $\kappa$, without considering the underlying model: such a feature would greatly simplify the study of 
the phenomenology of an additional light scalar.

\section*{Acknowledgements}
We thank the CMS $H\to\gamma\gamma$ group in Lyon: in particular Suzanne Gascon-Shotkin, Morgan Lethuillier, Louis Sgandurra 
and Fan Jaiwei for their support and for discussions. We also thank Nazila Mahmoudi for her help in using \texttt{SuperIso} and for 
providing the routine for $\Delta M_d$. GDLR thanks Cyril Hugonie for his support on technical questions related to 
\texttt{NMSSMTools}.  AD is partially supported by Institut Universitaire de France. We also acknowledge partial support from 
the Labex-LIO (Lyon Institute of Origins) under grant ANR-10-LABX-66 and FRAMA (FR3127, F\'ed\'eration de Recherche `Andr\'e 
Marie Amp\`ere"). 

\providecommand{\href}[2]{#2}\begingroup\raggedright\endgroup

\end{document}